\documentclass[aps,prbbib,twocolumn,epsf]{revtex4}

\usepackage{graphicx}
\usepackage[usenames]{color}

\begin{document}
\draft

\title{Depletion region surface effects in electron beam induced current measurements}
\author{Paul M. Haney$^1$, Heayoung P. Yoon$^{2}$, Benoit Gaury$^{1,3}$, Nikolai B. Zhitenev$^1$}

\affiliation{1.  Center for Nanoscale Science and Technology, National Institute of Standards and Technology, Gaithersburg, MD 20899, USA \\
2. Department of Electrical and Computer Engineering, University of Utah, Salt Lake City, UT 84112, USA\\
3.  Maryland NanoCenter, University of Maryland, College Park, MD 20742, USA
}
\begin{abstract}
Electron beam induced current (EBIC) is a powerful characterization technique
which offers the high spatial resolution needed to study polycrystalline solar
cells.  Current models of EBIC assume that excitations in the $p$-$n$ junction depletion region result in perfect charge collection efficiency.  However we find that in CdTe and Si samples prepared by focused ion beam (FIB) milling, there is a reduced and nonuniform EBIC lineshape for excitations in the depletion region.  Motivated by this, we present a model of the EBIC response for excitations in the depletion region which includes the effects of surface recombination from both charge-neutral and charged surfaces.   For neutral surfaces we present a simple analytical formula which describes the
numerical data well, while the charged surface response depends qualitatively on
the location of the surface Fermi level relative to the bulk Fermi level.  We find the experimental data on FIB-prepared Si solar cells is most consistent with a charged surface, and discuss the implications for EBIC experiments on polycrystalline materials.
\end{abstract}

\maketitle

\section{Introduction}

\begin{figure}[t]
\includegraphics[width=0.45\textwidth]{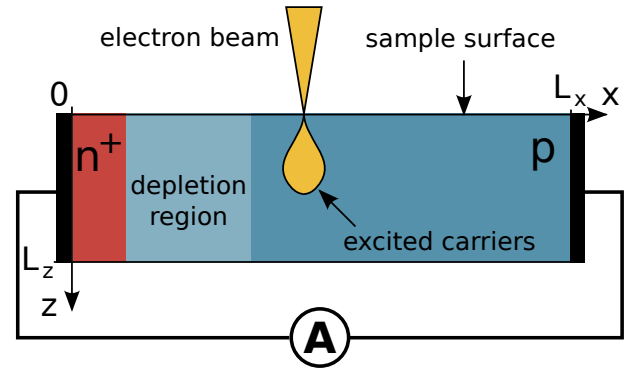}
\caption{Schematic of cross-sectional EBIC experiment performed on a $n^+$-$p$
junction.  The black layers on left and right sides of the device correspond to the $n$ and $p$ contacts, respectively.}\label{fig:1}
\end{figure}

Polycrystalline photovoltaic materials such as CdTe exhibit high power
conversion efficiency in spite of their high defect density~\cite{kumar}.  Grain
boundaries are a primary source of defects in these materials.  Understanding
the role of grain boundaries in the photovoltaic performance of these materials
requires quantitative information about the electronic properties at the length
scale of individual grains (typically $1~{\rm \mu m}$).  A measurement technique
which offers this spatial resolution is electron beam induced current (EBIC).
Fig.~\ref{fig:1} shows a schematic of an EBIC experiment: a beam of high energy
electrons generates electron-hole pairs in proximity to an exposed surface.  A
fraction $\eta$ of these electron-hole pairs are collected is the contacts and measured
as electrical current~\cite{Hanoka}, while the remaining fraction $1-\eta$ undergoes recombination.  We refer to $\eta$ as the EBIC collection efficiency.  EBIC has been used as a diagnostic tool for measuring
material properties such as the minority carrier diffusion length and
surface recombination velocity~\cite{wu,roosbroek,donolato}. The small
excitation volume associated with the electron beam is in close
proximity to the exposed surface, so the high spatial resolution of EBIC is
necessarily accompanied by strong surface effects on the measured signal.

Quantitative models of EBIC in single crystal materials are well established by Van Roosenbrook \cite{roosbroek}, Donolato \cite{donolato,donolato_EBIC2}, and Berz {\it et al.} \cite{berz}, among others, who obtained solutions for the EBIC collection efficiency versus excitation position in the cross-sectional geometry, including the effects of surface recombination, while Ioannou {\it et al.} \cite{ioannou} and Luke \cite{luke_planar} provide solutions and analysis for a planar contact geometry.  All previous EBIC models apply for excitations in the neutral region, and enable the extraction of the minority carrier diffusion length, and the ratio of the surface recombination velocity to the minority carrier diffusivity.  All of these models further assume that minority carriers in the $p$-$n$ junction depletion region are collected with $100~\%$ efficiency \cite{footnote1}.  As we discuss next, this assumption is not always satisfied in EBIC experiments.

For CdTe, the EBIC response observed in several recent works deviates from the expected behavior based on previous EBIC models.  For example, the depletion width of the $p$-$n$ junction is on the order of 1 ${\rm \mu m}$,  and occupies a substantial fraction of the absorber thickness (which is typically on the order of 3 ${\rm \mu m}$).  Previous EBIC models assume $100~\%$ collection from the depletion region, however experiments show a nonuniform EBIC response throughout the absorber which is well below $100~\%$ \cite{yoon,li,zywitzki}.  In addition, the EBIC response near grain boundaries does not conform to existing models.  These models relate the grain boundary recombination velocity to the {\it reduction} of EBIC collection efficiency at the grain boundary core \cite{donolato_GB1, donolato_GB2, corkish,chen1}.  However in CdTe the EBIC collection efficiency is {\it maximized} at the grain boundary core \cite{yoon,li,zywitzki}.

Motivated by these observations, we present an extension of previous EBIC models to consider the response for excitations in the depletion region.  We find analytical expressions and present numerical results for the EBIC collection efficiency which includes recombination from neutral and charged surfaces.  To validate the relevance of the model, we also present experimental EBIC data on Si solar cells prepared by cleaving and focused ion beam (FIB) milling.  We find that the FIB-prepared sample exhibits a suppressed maximum EBIC collection efficiency, and that the behavior is most consistent with the model of EBIC response in the presence of a charged surface.

The paper is organized as follows: in Sec.~\ref{sec:expt} we present
experimental results of the cross-sectional EBIC response of Si solar cells where the sample surfaces are prepared with cleaving and FIB milling.  In Sec.~\ref{sec:neutral}, we present 2-d numerical simulation results and corresponding analytical
expressions for the EBIC efficiency in the depletion region which includes
recombination at a neutral surface.  In Sec.~\ref{sec:charged}, we
extend the model to consider surfaces with charged defect levels.  We end with a
comparison of the model predictions to the experimental data, which shows that
the charged surface model is most consistent with the experimental observations.

\section{Experiment}\label{sec:expt}

We start with the experimental data which demonstrates the need for a model of EBIC response in the depletion region.  Two sets of EBIC data were obtained from a commercially available, single crystalline Si solar panel (power conversion efficiency of $20~\%$).  The absorber thickness is approximately $300~{\rm \mu m}$.
A conventional capacitance-voltage measurement yields a doping density of $10^{15}~{\rm cm^{-3}}$ in the Si absorber layer. We use the native contacts on the $p$-type and $n$-type layers of the solar cell for the EBIC measurement. Following the
cleaving process, we carried out an additional ion milling
process for one of the samples. A focused Ga ion beam
at 30 keV (beam current of 2.5 nA) was irradiated on
top of the sample, and the exposed area was etched
away. This type of FIB process is often used to create smooth
sections of photovoltaic devices prior to their local
electrical and optical characterizations. The
cross-sections of the cleaved and the FIB prepared Si
samples were very smooth \cite{surface}, minimizing the artifacts
arising from the surface roughness. A series of EBIC
data was obtained at different electron beam energies to
examine the depth dependence of the EBIC response inside the depletion
region and away from the $p$-$n$ junction.

\begin{figure}[h!]
\begin{center}
\vskip 0.2 cm
\includegraphics[width=8cm]{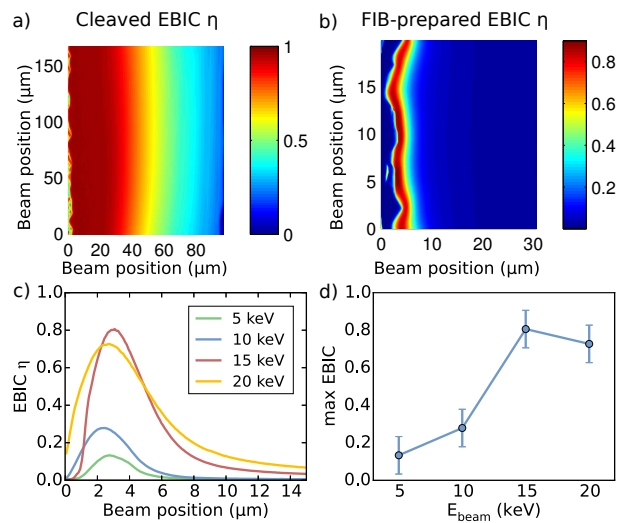}
\vskip 0.2 cm \caption{(a) EBIC collection efficiency map for the cleaved sample at beam energy of 15 keV.  The $n$-contact is at the left-hand side.  (b) The same data for the FIB-prepared sample.  (c) EBIC collection efficiency versus beam position for FIB'd Si samples, for beam energies of $(5,10,15,20)~{\rm keV}$, (d) the corresponding maximum collection efficiency versus beam energy.}\label{fig:fit1}
\end{center}
\end{figure}

To determine the absolute EBIC collection efficiency $\eta$, we estimate the generation rate of electron-hole pairs $G_{\rm tot}$ using the empirical relation \cite{wu}:
\begin{eqnarray}
G_{\rm tot} = \left(1-b\right)\frac{\left(I_{\rm beam}/q\right)\times \left(E_{\rm beam}/E_0\right)}{2.59 \times \left(E_g/E_0\right) + 0.71}, \label{eq:G}
\end{eqnarray}
 where $I_{\rm beam}$ is the electron beam current, $q$ is the electron charge, $E_{\rm beam}$ is the beam energy, $E_g$ is the material bandgap, $E_0=1~{\rm eV}$, and $b$ is the backscattering coefficient, corresponding to the fraction of reflected energy. $b$ is determined by Monte Carlo calculations \cite{srim}; we find $b$ is approximately 0.15 and varies slightly with beam energy.  The electron beam current varies with the beam energy, and is in the range of 200 pA to 500 pA.  The EBIC collection efficiency is the ratio of the measured current to $G_{\rm tot}$.

We estimate 10~\% uncertainty in the measured EBIC efficiency $\eta$ (all uncertainties are reported as one standard deviation).  The dominant sources of uncertainty are from the beam current, the backscattering coefficient, which is computed for pure, uncontaminated Si, and the reliability of the empirical relation we use Eq.~(\ref{eq:G}) to estimate the rate of electron-hole pair generation.

EBIC collection efficiency maps are shown in Fig.~\ref{fig:fit1}(a) and (b) for the cleaved and FIB-prepared samples, respectively, for an electron beam energy of 15 keV.  For the FIB-prepared sample, the maximum EBIC efficiency is reduced and the collection efficiency decays much more rapidly as compared to the cleaved sample.  Plots of the EBIC collection efficiency versus excitation position for several electron beam energies are shown in Fig.~\ref{fig:fit1}(c) for the FIB-prepared sample.  The maximum collection efficiency is much less than 100~\% and only recovers to 0.85 for a beam energy of 20 keV (see Fig.~\ref{fig:fit1}(d)).  In contrast, for cleaved samples we find that the maximum EBIC collection efficiency is 100~\% (within the experimental uncertainty) for all electron beam energies.  The reduced EBIC collection efficiency for FIB-prepared samples is therefore due to the sample surface preparation, indicating the importance of including surface effects in EBIC models.

We have checked that the electron-hole pair excitation rate is low enough to avoid ``high-injection" nonlinearities in the EBIC response \cite{Haney}.  The excitation bulb size at energies less than 10 keV is less than the depletion width of the $p$-$n$ junction (which is approximately $1~\mu m$), so that the reduced efficiency is not the result of an extended excitation profile \cite{luke_genvol}.  As discussed in the introduction, the observation of less than 100~\% collection efficiency for excitations in the depletion region violates assumptions of previous EBIC models, motivating the models we present in the next sections.

\section{Recombination from neutral surfaces}\label{sec:neutral}

We consider the collection efficiency for an excitation in the depletion region
for a 2-d model, which includes recombination from the exposed surface. In this
section the surface is assumed to be uncharged.

\bigskip
Fig.~\ref{fig:1} shows our geometry and coordinate system. For the
\mbox{2-d} finite difference simulations, we solve the electron and hole continuity
equations together with the Poisson equation. The depth of the
system away from the surface ($L_z$) is large enough to ensure that the lower boundary
does not affect the results. We use a highly non-uniform mesh,
with grid spacing as small as $1~{\rm nm}$ near the surface in order to properly
resolve the generation profile and density gradients there.

We assume selective contacts, such that the contact surface recombination is
infinite (zero) for majority (minority) carriers.  The exposed surface
recombination is included by adding an extra recombination term at the surface,
\begin{eqnarray}
R_{\rm surf} = \delta\left(z\right) \times S \frac{n_s p_s-n_i^2}{n_s + p_s +
n_{\rm surf} + p_{\rm surf}},
\label{eq:surfR}
\end{eqnarray}
where the surface recombination velocity $S$ parametrizes the magnitude of the
surface recombination. $n_s$ ($p_s$) is the electron (hole) density evaluated
at the surface, and $n_{\rm surf}$, $p_{\rm surf}$ are
\begin{eqnarray}
n_{\rm surf} &=& N_C\exp\left(\frac{E_{\rm surf}-E_C}{k_{\rm B}T}\right)\\
p_{\rm surf} &=& N_V\exp\left(\frac{E_V-E_{\rm surf}}{k_{\rm B}T}\right),
\end{eqnarray}
where $E_{\rm surf}$ is the surface defect energy level, $E_V$ ($E_C$) is the valence
(conduction) band edge energy, $N_C$ ($N_V$) is the conduction and valence
effective band density of states, $k_{\rm B}$ is the Boltzmann constant, and $T$ is the temperature.  For the neutral surface we take $E_{\rm surf}$ to be at midgap.

The generation profile for a beam positioned at $x_B$ is taken to be Gaussian:
\begin{eqnarray}
G(x,z) = \frac{A}{2\pi\sigma^2}
\exp\left[-\frac{\left(x-x_B\right)^2+\left(z-z_B\right)^2}{2\sigma^2}\right],
\label{eq:Gxz}
\end{eqnarray}
where $z_B=0.3~R_B$, $\sigma = R_B/\sqrt{15}$~\cite{donolato}, and the excitation bulb size $R_B$ is
given in the following empirical relation \cite{grun}:
\begin{eqnarray}
R_B=\frac{0.043\times R_0}{\left(\rho/\rho_0\right)}\left(E_{\rm beam}/E_0'\right)^{1.75}\label{eq:GR}
\end{eqnarray}
where $\rho$ is the material mass density, $\rho_0=1~{\rm g/cm^3}$, $E_0'=1~{\rm keV}$, and $R_0=1~{\rm \mu m}$.

The units of the total generation in the
2-d model are ${\rm s^{-1}\cdot m^{-1}}$.  To convert the experimental
generation rate $G_{\rm tot}$ (which has units ${\rm s^{-1}}$) into a 2-dimensional
generation rate, we divide $G_{\rm tot}$ by the diffusion length $L_D$.  In Eq.~(\ref{eq:Gxz}), $A$ is chosen such that the spatial integral of $G(x,z)$ in the sample is equal to $G_{\rm tot}/L_D$.  Table~\ref{params} gives a list of the parameters used in the numerical calculations.

The beam current used in the simulation is less than the experimental value.  We find that using the experimental value in a 2-d simulation results in high injection effects and screening of the $p$-$n$ junction.  However, the experiment is not in the high injection regime, as evidenced by the lack of beam current dependence, the undistorted EBIC lineshapes \cite{Nichterwitz}, and the estimate for the critical beam current for high injection given in Ref. \onlinecite{Haney}.  The reason for this discrepancy is that the onset of high injection effects depends strongly on system dimensionality \cite{Haney}.   We therefore reduce the electron beam current of the model to prevent high injection effects.

\begin{table}
\setlength{\tabcolsep}{1cm}
\begin{tabular}{ll}
  \hline \hline
  Parameter & Value \\ \hline
  $N_C=N_V$ & $10^{19}~{\rm cm^{-3}}$ \\
  $E_g$ & $1.1~{\rm eV}$ \\
  $N_A$ & $10^{14}~{\rm cm^{-3}}$ to $10^{15}~{\rm cm^{-3}}$ \\
  $N_D$ & $10^{18}~{\rm cm^{-3}}$ \\
  $\mu_e=\mu_h$ & $10~{\rm cm^2/\left(V\cdot s\right)}$ \\
  $\epsilon$ & $11$ \\
  $\tau_{\rm bulk}$ & $10^{-6}~{\rm s}$ \\
  $L_x$ & $30~{\rm \mu m}$ \\
  $L_z$ & $56~{\rm \mu m}$ \\
  $I_{\rm beam}$ & $ 1~{\rm pA}$ \\
  $S$ & $10^6 ~{\rm cm/s}$ \\
  $N_{\rm surf}$ & $10^{11} ~{\rm cm^{-2}}$ \\
  \hline \hline
\end{tabular}
\caption{List of default parameters for numerical simulations.\label{params}}
\end{table}

\begin{figure}[h!]
\includegraphics[width=0.48\textwidth]{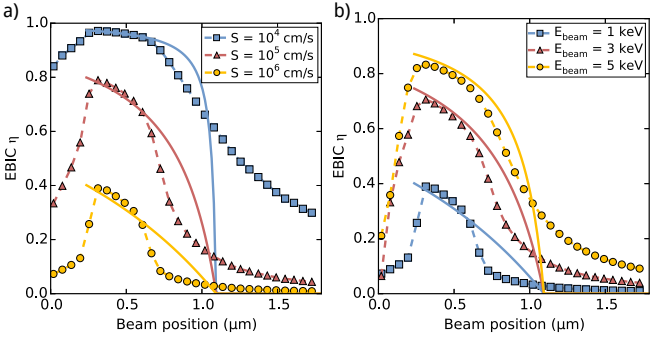}
\caption{(a)  Simulated EBIC linescans for a series of surface recombination velocities, with fixed beam
energy of $1~{\rm keV}$.  (b) Simulated EBIC linescans for a series of electron
beam energies, with a fixed $S=10^6~{\rm cm/s}$.  For both plots, circles indicate numerical
results, while solid lines show the prediction of Eq.~(\ref{eq:surf2_R}).}\label{fig:neutral}
\end{figure}


\bigskip
Fig.~\ref{fig:neutral}(a) shows the computed EBIC efficiency as a function of beam
position for a fixed beam energy of $1~{\rm keV}$ and three values of $S$.  The
maximum efficiency is strongly reduced for increasing $S$.
We propose the following expression to describe the EBIC efficiency when the excitation is localized near the surface:
\begin{eqnarray}
\eta(E_{\rm beam}\rightarrow0) = \frac{\mu \mathcal{E}}{\mu \mathcal{E} + S/2}, \label{eq:surf1}
\end{eqnarray}
where $\mu$ is the smaller of the electron and hole mobility, and $\mathcal{E}$ is the (position-dependent) electric field.  Eq.~(\ref{eq:surf1}) has a simple physical interpretation. The collection
efficiency is determined from the competition between drift current and surface
recombination.  The total collection rate from drift current is $A_x
\times\left(n \mu \mathcal{E}\right)$, where $n$ is the carrier density at the excitation
point, and $A_x$ is the area of the excitation normal to the direction of the
drift current.  The total recombination rate is $A_z \times \left( n
S/2\right)$, where $A_z$ is the area of the excitation normal to the surface.
The factor of $1/2$ multiplying $S$ follows from the assumption that $n\approx p$ near the excitation, which reduces the recombination rate at the surface by $1/2$ (see Eq.~(\ref{eq:surfR})). This assumption is easily satisfied for excitations in the depletion region. For the small, symmetrical excitations considered here, $A_x \approx A_z$.  Forming the ratio of the collection rate to
the sum of the collection and recombination rates leads to Eq.~(\ref{eq:surf1}).


Larger beam energies generate electron-hole pairs further from the surface,
reducing its effect on the EBIC collection efficiency.  This is demonstrated in the
simulation data shown in Fig.~\ref{fig:neutral}(b), where the maximum EBIC
efficiency increases with increasing beam energy.  We find that the following
expression accounts for the dependence of the surface recombination on the
excitation depth $z_B$:
\begin{eqnarray}
\eta &=& 1-\left(\frac{S/2}{\mu \mathcal{E} + S/2}\right)  \left(\frac{D/z_B}{\mu
\mathcal{E}+D/z_B}\right).
\label{eq:surf2_R}
\end{eqnarray}
In the above, the first factor in parentheses is the recombination probability
for a charge located {\it at} the surface.  This is given by the ratio of the
recombination velocity to the sum of drift and recombination velocities (see
discussion below Eq.~(\ref{eq:surf1})). The second factor in
Eq.~(\ref{eq:surf2_R}) represents the probability a charge located at $z_B$
below the surface will diffuse to the surface.

The solid lines of Fig.~\ref{fig:neutral} show the estimated EBIC collection efficiency given by Eq.~(\ref{eq:surf2_R}), using the standard form of the $p$-$n$ junction electric field $\mathcal{E}(x)$.  We observe good agreement deep within the depletion region (for beam positions between 0.3 and 0.7 ${\rm \mu m}$) between the analytical and numerical models.  In this region the assumption $n\approx p$ is satisfied.  For beam positions outside of this interval, a disparity between the numerical and analytical models is due to a violation of the $n\approx p$ assumption.  The maximum EBIC signal occurs deep within the junction so that Eq.~(\ref{eq:surf2_R}) reliably predicts the maximum EBIC collection efficiency.

\begin{figure}[h!]
\includegraphics[width=0.46\textwidth]{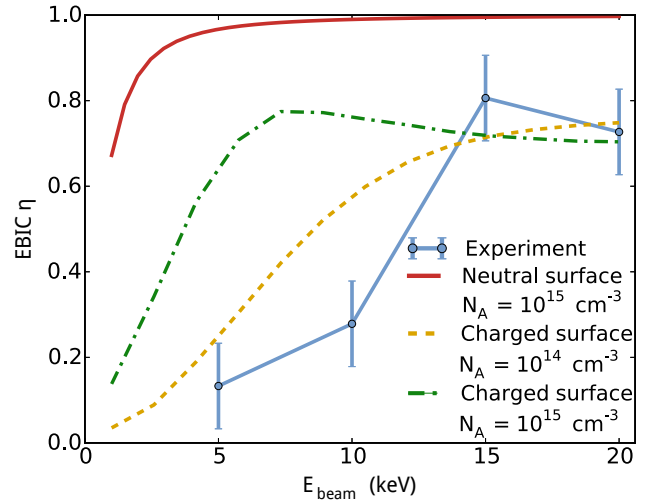}
\caption{Maximum EBIC efficiency versus beam energy for the neutral and charged surface 2-d models, and the experimental data.  In the models, $S=10^6~{\rm cm/s}$.  For the charged surface model, $E_{\rm surf}=0,~N_{\rm surf}=10^{11}~{\rm cm^{-2}}$, and the results for two doping densities are shown.}\label{fig:comparison2d}
\end{figure}

The predicted maximum EBIC efficiency versus electron beam energy is shown in Fig.~\ref{fig:comparison2d} as a solid red line.  The surface recombination has a minor impact on EBIC efficiency for electron beam energies
greater than $5~{\rm keV}$, corresponding to excitation bulb sizes greater than
$300~{\rm nm}$.  This can be understood from Eq.~(\ref{eq:surf2_R}), which shows that surface recombination becomes negligible for $z_B \gg D/\left(\mu \mathcal{E}\right)$. For relevant material parameters, this corresponds to $z_B \gg 10~{\rm nm}$.

The experimental data also shown in Fig.~\ref{fig:comparison2d} indicate that surface effects are present at much larger electron beam energies than those predicted by the neutral surface model.  While our numerical and analytical work provide a full picture of the neutral surface model, the experimental data are at odds with these results, indicating that other factors not included in this model play a key role in the experiment.  In the next
section, we consider electric fields at the surface induced by charged surface
defects.

\section{Recombination from charged surfaces}\label{sec:charged}

We next consider charged surface defect states.  The surface charge density $\rho_{\rm surf}$ is
determined by the occupancy $f_{\rm surf}$ of defect energy level~\cite{rau}:
\begin{eqnarray}
\rho_{\rm surf}=q \frac{N_{\rm surf}}{2}\left(1-2f_{\rm surf}\right),\label{eq:rhosurf}
\end{eqnarray}
where $N_{\rm surf}$ is the 2-d density of surface defect states and $f_{\rm surf}$ is given by
\begin{eqnarray}
f_{\rm surf} &=& \frac{n_s + p_{\rm surf}}{n_s + p_s + n_{\rm surf} + p_{\rm surf}}.
\end{eqnarray}
\begin{figure}[h!]
\includegraphics[width=0.46\textwidth]{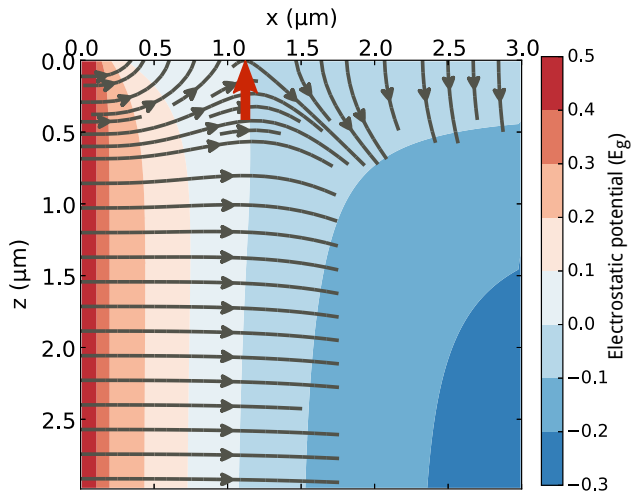}
\caption{Numerically computed electrostatic potential in units of the band gap energy  as a function
of 2-d position in the depletion region.  $z=0$ corresponds to the surface, at which the
potential is pinned to the midgap surface defect level (white).  The gray lines
are electric field lines showing the direction of the local electric field.  The red arrow indicates the position $x_0$ at which the surface energy level equals the bulk $p$-$n$ junction Fermi level.}
\label{fig:charged_electrostatics}
\end{figure}
We add Eq.~(\ref{eq:rhosurf}) to the drift-diffusion-Poisson numerical model.
Fig.~\ref{fig:charged_electrostatics} shows the computed equilibrium electrostatic
potential for a surface defect energy level at the midgap and $\rho_{\rm
surf}=10^{11}~\rm cm^{-2}$.  As expected, the Fermi level is pinned to the
defect energy level at the surface, which leads to dramatic changes
in the electrostatic potential of the $p$-$n$ junction.

An important transition occurs where the bulk $p$-$n$ junction electrochemical potential is equal
to the surface defect energy level - this is shown as a red arrow in Fig.~\ref{fig:charged_electrostatics}, and we denote this position $x_0$.
For positions greater than $x_0$, the surface electrostatic field is directed
{\it away} from the surface, while for positions less than $x_0$, the surface
field is directed {\it toward} the surface.  The surface field generally changes
directions within the $p$-$n$ junction for any surface defect level which is
positioned between the Fermi levels of the $n$ and $p$ type semiconductors.  The
surface field drives electrons towards the surface for $x>x_0$, and drives holes
to the surface for $x<x_0$.  We emphasize that the position of $x_0$ depends strongly on the surface defect energy level.

\begin{figure}[h!]
\includegraphics[width=0.46\textwidth]{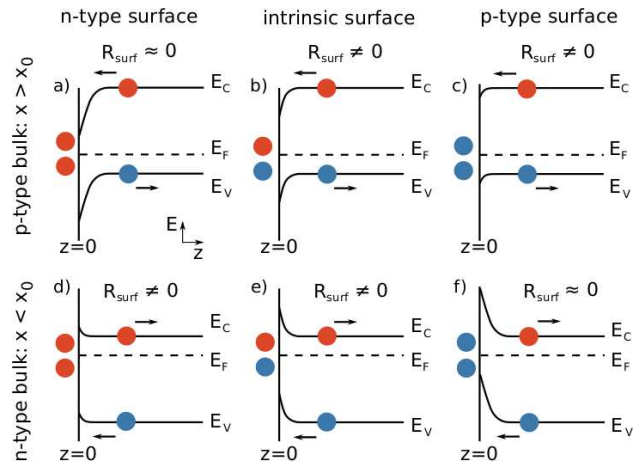}
\caption{Depiction of band diagrams versus position away from the surface $z$, at a position $x>x_0$ (upper row) where the bulk is $p$-type, and at a position $x<x_0$ (lower row) where the bulk is $n$-type.  The surface may be $n$-type, intrinsic, or $p$-type.  Depending on the combination of surface and bulk properties, recombination is enhanced or suppressed at the surface, as shown in the various cases.  Electrons (holes) are represented with red (blue) dots, and arrows indicate the drift direction of each carrier type due to the surface electric field.}
\label{fig:bands}
\end{figure}

We start with a qualitative discussion of the system behavior. Far enough from the surface, there is drift along the bulk $p$-$n$ junction. Holes (electrons) drift parallel (anti-parallel) to the field lines of Fig. \ref{fig:charged_electrostatics}. Closer to the surface, carriers drift towards/away from the surface, which is the dominant recombination center.  The minority carrier type at the surface depends on the surface defect energy level. If the carriers driven to the surface by the surface electric field are minority carriers there, then recombination occurs and the EBIC signal is decreased (this corresponds to Figs. \ref{fig:bands}(c) and (d)). On the other hand, if the carriers driven to the surface are majority carriers there, then recombination does not take place (this corresponds to Figs. \ref{fig:bands}(a) and (f)). Due to Fermi level pinning at the surface, there is no potential gradient along it so that majority carriers undergo simple diffusion to the contacts along the surface, and the EBIC collection efficiency is $\approx100~\%$. The various cases of surface and bulk types and the resultant surface recombination $R_{\rm surf}$ are shown in Fig. \ref{fig:bands}.


\begin{figure}[h!]
\includegraphics[width=0.46\textwidth]{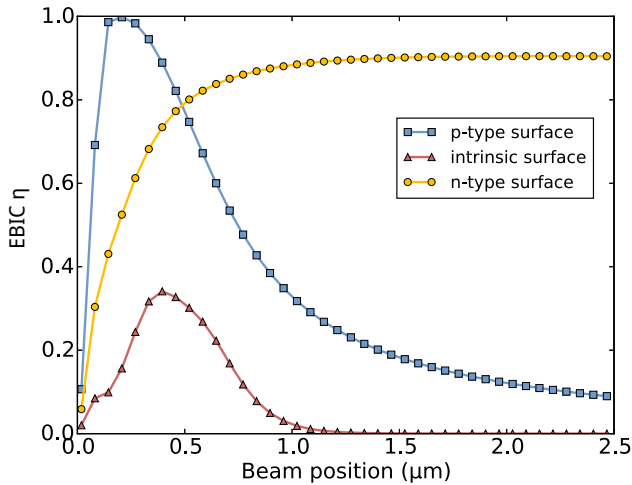}
\caption{EBIC efficiency as a function of beam
position for three values of the surface defect level. $E_{\rm surf}=-0.38~\rm eV$
corresponds to a
$p$-type surface (red), $E_{\rm surf}=0~\rm eV$ to an intrinsic surface (black), and
$E_{\rm surf}=0.38~\rm eV$ to an $n$-type surface (blue).}\label{fig:charged}
\end{figure}

Fig.~\ref{fig:charged} shows that the impact of the surface field on the EBIC efficiency
depends on the surface type ($n$, $p$, or intrinsic).  We first
consider a $p$-type surface.  The collection efficiency is maximized for
excitation positions less than $x_0$ ({\it i.e.} where the field is directed toward the surface).  In this case the surface field drives
holes toward the surface, where they are majority carriers and therefore undergo
no recombination.  On the other hand, for excitation positions greater than
$x_0$ ({\it i.e.} where the field is directed away from the surface) the surface field drives electrons to the $p$-type surface.  Because
electrons are minority carriers there, they undergo recombination and the the EBIC collection efficiency is reduced.

For an $n$-type surface, the situation is reversed; for excitation positions less than $x_0$, the surface field drives holes to the $n$-type surface, where they recombine, while for excitations positioned beyond $x_0$, the surface field drives electrons to the $n$-type surface, where they avoid recombination.  This results in an ``inverted" EBIC signal in which the collection efficiency is maximized away from the $p$-$n$ junction.  This type of EBIC signal has been previously observed~\cite{hungerford,bailey}, and attributed to surface passivation via the electrostatic surface field, as described here.

Finally, for an intrinsic surface, there is no clearly defined majority/minority carrier and both electrons and holes undergo recombination at the surface.  In this case, the collection efficiency is maximized at the excitation position $x=x_0$; at this position the magnitude of the surface field is minimized, so carries aren't driven to the surface here.  In this case, the maximum EBIC efficiency is significantly less than 1, consistent with the experimental observations.

The length scale in the $z$-direction over which the surface influences the EBIC collection efficiency is set by the depletion width of the surface, which depends on the surface defect level and bulk doping.  This length scale is generally much larger than that of the neutral surface recombination.  For example, a midgap surface defect level and $p$-type doping of $10^{14}~{\rm cm^{-3}}$ results in a surface depletion width of $1.5~{\rm \mu m}$.

\begin{figure}[h!]
\includegraphics[width=0.48\textwidth]{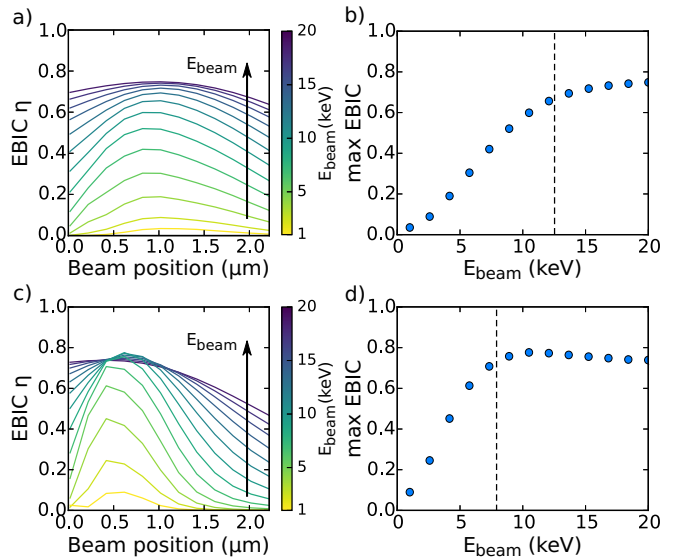}
\caption{(a) EBIC lineshapes as a function of beam positions for values of
$E_{\rm beam}$ equally spaced between 1 and 20 ${\rm keV}$. The arrow indicates the direction of increasing electron beam energy.  Bulk doping is
$N_A=10^{14}~{\rm cm^{-3}}$, defect energy level is at midgap, and $N_{\rm
surf}=10^{11}~{\rm cm^{-2}}$. (b) Maximum EBIC efficiency as a function of
$E_{\rm beam}$.  The vertical line corresponds to the beam energy for which the
associated excitation bulb size is equal to the surface depletion width.  (c), (d) Same data as (a) and (b) for an increased value of doping, $N_A=5\times10^{14}~{\rm cm^{-3}}$.}\label{fig:doping}
\end{figure}


Fig.~\ref{fig:doping}(a) shows the simulated EBIC lineshape for several electron beam energies for a $p$ doping level of $N_A=10^{14}~{\rm cm^{-3}}$ and midgap surface defect level.  As expected, the surface effect is diminished for larger beam energies (larger excitation bulb sizes).  The dotted vertical line in Fig.~\ref{fig:doping}(b) indicates the surface depletion width, and the maximum EBIC efficiency begins to saturate for excitation bulbs sizes which exceed this.  Fig.~\ref{fig:doping}(c) and (d) show the same data for an increased doping level ($N_A=5\times 10^{14}~{\rm cm^{-3}}$) and smaller associated surface depletion width.  As expected, the surface influence is diminished at lower electron beam energies in this case.

Fig.~\ref{fig:comparison2d} shows the maximum EBIC efficiency versus electron beam energy observed experimentally, together with the predictions of the neutral and charge surface models.  It is clear that the charged surface model is more consistent with the data.  For a doping density of $10^{15}~{\rm cm^{-3}}$, the EBIC collection efficiency increases with electron beam energy more rapidly than observed experimentally.  However, the value of EBIC efficiency at high electron beam energies remains well below 1.  We consider this to be a signature of surface recombination from charged surfaces.  Reducing the doping density to $10^{14}~{\rm cm^{-3}}$ in the model yields a better fit to experimental data.  We also find that to obtain the best fit to experimental data, the value of the mobility $\mu$ must be substantially lower than the expected Si mobility (we use $\mu=10~{\rm cm^2/\left(V\cdot s\right)}$, see Table 1).  Since most of the carrier transport occurs near the sample surface, it is plausible that for the FIB-prepared sample the mobility of carriers near the surface is substantially less than the bulk mobility, due to disorder introduced in the sample surface preparation.

We can not rule out the possibility that other factors are responsible for the experimental observations.  However we note that other surface sensitive studies applied to $p$-$n$ junctions, such as scanning Kelvin probe microscopy, have relied on surface fields in their models in order to qualitatively describe their experimental results \cite{rosenwaks}.  We have also studied the possibility of bulk recombination in the depletion region as a mechanism for the reduced EBIC efficiency, but this recombination mechanism would apply in an operational device, which would be inconsistent with the high short circuit current density of the devices.  For these reasons, we believe that charged surface defect levels are the most likely explanation for the reduced EBIC efficiency observed in these materials.

If indeed surface fields significantly impact an EBIC measurement, then conclusions drawn from these measurements should be framed in the context of a surface field.  For example, the observations that excitations at grain boundaries result in higher EBIC collection efficiency than excitations in grain interiors may be used to conclude that grain boundaries are not recombination centers {\it in the presence of surface fields} \cite{li,yoon}.  The extent to which this qualification modifies the conclusions about grain boundaries in the bulk is not clear {\it a priori}, and we leave this as a subject for future study.  We also note that the presence and magnitude of surface fields likely depends on sample preparation methods.  We suggest that measuring the maximum EBIC efficiency as a function of electron beam energy is one way to determine if surface fields are present in a particular sample, but acknowledge that a definitive conclusion about the presence of surface fields is generally difficult.

\section{Conclusion}

In summary, we've developed a model of the EBIC response to excitations in the depletion region which includes surface recombination from neutral and charged surfaces.  This model is motivated by observations that, under certain experimental conditions, the maximum EBIC collection efficiency is substantially less than 100~\% and varies throughout the depletion region.  A comparison between the model and Si samples offers clear indication of the role of surface effects in EBIC experiments, and how they depend on sample preparation methods.  A particularly useful and common application of low energy, high resolution EBIC is probing the characteristics of polycrystalline photovoltaics.  However the interpretation of the EBIC response in these materials is complicated by the presence of grain boundaries.  Using a simpler material such as Si offers a cleaner interpretation of surface effects and development of surface models.  Ultimately these surface models can be applied to more complex materials, enabling a determination of properties of grain boundary and subsurface interfaces.  Moreover, models of the EBIC response at the sample surface can be readily applied to internal surfaces (e.g. grain boundaries), so that the formulas and qualitative descriptions given here may be applied to studies of grain boundaries as well.

\section*{Acknowledgment}
B. Gaury acknowledges support under the Cooperative Research Agreement between the University of Maryland and the National Institute of Standards and Technology Center for Nanoscale Science and Technology, Award 70NANB10H193, through the University of Maryland.  H. Yoon also acknowledges the support by the NSF MRSEC program at the University of Utah under grant \# DMR 1121252.

\begin{appendix}

\end{appendix}


\begin{thebibliography}{1}

\bibitem{kumar}
S. G. Kumar, K. S. R. K. Rao, Energy and Environmental Science {\bf 7}, 45 (2014).

\bibitem{Hanoka}
J.I. Hanoka, R.O. Bell, Annu. Rev. Mater. Sci. {\bf 11}, 353 (1981).

\bibitem{wu} C. J. Wu and D. B. Wittry, J. App. Phys., {\bf 49}, 2827, (1978).

\bibitem{roosbroek}
W. can Roosbroeck, J Appl. Phys. {\bf 26}, 380 (1955).

\bibitem{donolato}
C. Donolato,  Solid-State Elect. {\bf 25}, 1077 (1982).


\bibitem{donolato_EBIC2}
C. Donolato, App. Phys. Lett. {\bf 43}, 120 (1983).

\bibitem{berz}
F. Berz and H. K. Kuiken, Sol. St. Elect. {\bf 19} 437 (1976).

\bibitem{ioannou}
D. E. Ioannou and C. A. Dimitriadis, IEEE Trans. on Elect. Dev. {\bf 29}, 445 (1982).


\bibitem{luke_planar}
K. L. Luke, J. App. Phys. {\bf 80}, 5775 (1996).


\bibitem{footnote1}
The assumption that all carriers within the depletion region are collected is contained in the boundary condition that the minority carrier density vanishes at the depletion region edge.


\bibitem{li}
C. Li, Y. Wu, J. Poplawsky, T. J. Pennycook, N. Paudel, W. Yin, S. J. Haigh, M. P. Oxley, A. R. Lupini, M. Al-Jassim, S. J. Pennycook, and Y. Yan, Phys. Rev. Lett. {\bf 112}, 156103 (2014).

\bibitem{yoon} H. P. Yoon, P. M. Haney, D. Ruzmetov, H. Xua, M. S. Leite,
B. H. Hamadani, A. Talin, and N. B. Zhitenev, Sol. Energy Mat. and Solar Cells {\bf 117},  499, (2013).

\bibitem{zywitzki}
O. Zywitzki, T. Modes, H. Morgner, C. Metzner, B. Siepchen, B. Sp\"{a}th, C.
Drost, V. Krishnakumar, and S. Frauenstein, J. App. Phys. {\bf 114}, 163518 (2013).


\bibitem{donolato_GB1}
C. Donolato, J. Appl. Phys. {\bf 54}, 1314 (1983).

\bibitem{donolato_GB2}
C. Donolato, Mat. Sc. and Eng. {\bf B24}, 61 (1994).

\bibitem{corkish}
R. Corkish, T. Puzzer, A. B. Sproul, and K. L. Luke, J. App. Phys {\bf 84}, 5473 (1998).

\bibitem{chen1}
J. Chen, T. Sekiguchi, D. Yang, F. Yin, K. Kido and S. Tsurekawa, J. App. Phys {\bf 96}, 5490 (2004).




\bibitem{surface}
H. Yoon, P. M. Haney, J. Schumacher, K. Siebein, Y. Yoon, N. B. Zhitenev, Microscopy and Microanalysis, {\bf 20}, 544 (2014).




\bibitem{srim}
J. F. Ziegler, M. D. Ziegler, and J. P. Biersack, Nucl. Inst. and Methods in Phys. Res. Sec. B: Beam Int. with Mat. and Atoms {\bf 268}, 1818 (2010).


\bibitem{Haney}
P. M. Haney, H. P. Yoon, P. Koirala, R. W. Collins, N. B. Zhitenev, Nanotech. {\bf 26}, 295401 (2015).

\bibitem{luke_genvol}
K. L. Luke, O. v. Roos, and L.-j. Cheng, J. App. Phys. {\bf 57}, 1978 (1985).

\bibitem{grun}
A. E. Gr\"{u}n, Zeitschrift f\"{u}r Naturforschung, {\bf 12a} (1957) 89.

\bibitem{Nichterwitz}
M. Nichterwitz and T. Unold, J. App. Phys. {\bf 144}, 134504 (2013).


\bibitem{rau}
K. Taretto and U. Rau, J. App. Phys. {\bf 103}, 094523 (2008).

\bibitem{bailey}
R. Hakimzadeh and S. G. Bailey, J. Appl. Phys. {\bf 74}, 1118 (1993).

\bibitem{hungerford}
G. A. Hungerford and D. B. Holt, {\it Institute of Physics Conference Series} {\bf 87}, 721 (1987).





\bibitem{rosenwaks}
S. Saraf and Y. Rosenwaks, Surf. Sci. {\bf 574}, L35 (2005).














\end{thebibliography}
\end{document}